\documentclass[sigplan,nonacm,10pt]{acmart}
\usepackage{microtype}
\usepackage{graphicx}
\usepackage{booktabs} 
\usepackage{soul}
\usepackage{color}
\usepackage{hyperref}
\usepackage{cleveref}
\usepackage{subcaption}
\usepackage{xspace}
\usepackage[export]{adjustbox} 
\usepackage{makecell}   
\usepackage{multirow}
\usepackage{tabularx}
\usepackage{url}
\usepackage{rotating}
\usepackage{array}
\usepackage{balance}
\usepackage{enumitem}
\usepackage{xcolor}
\usepackage[ruled,linesnumbered,nofillcomment]{algorithm2e}
\usepackage{ulem}
\usepackage{setspace}

\usepackage{fontawesome5}
\def\pname{XPerf\xspace}

\renewcommand\footnotetextcopyrightpermission[1]{}

\newlength{\ibleft}
\newlength{\ibright}

\newcommand{\insightbox}[1]{%
    \par\noindent
    {\setlength{\fboxsep}{0pt}%
    \fbox{%
        \hbox to \dimexpr\columnwidth-2\fboxrule\relax{%
            \hspace{\ibleft}%
            \parbox{\dimexpr\columnwidth-2\fboxrule-\ibleft-\ibright\relax}{%
                \begin{itemize}[leftmargin=5pt,labelsep=2.5pt,topsep=5pt,itemsep=0pt,parsep=0pt]
                    #1
                \end{itemize}%
            }%
            \hspace{\ibright}%
        }%
    }}%
}

\definecolor{darkcerulean}{rgb}{0.03, 0.27, 0.49}
\definecolor{codegreen}{rgb}{0,0.6,0}
\definecolor{codegray}{rgb}{0.5,0.5,0.5}
\definecolor{codepurple}{rgb}{0.58,0,0.82}
\definecolor{backcolour}{rgb}{0.95,0.95,0.92}
\definecolor{ballblue}{rgb}{0.13, 0.67, 0.8}
\definecolor{blue}{rgb}{0.0, 0.23, 0.84}
\definecolor{cobalt}{rgb}{0.0, 0.28, 0.67}
\definecolor{coolblack}{rgb}{0.0, 0.18, 0.39}
\definecolor{darkcerulean}{rgb}{0.03, 0.27, 0.49}
\definecolor{readgreen}{rgb}{0, 0.398, 0.199}
\definecolor{writepurple}{rgb}{0.906, 0.5195, 0.906}
\definecolor{compactpurple}{rgb}{0.48, 0, 0.5078}
\definecolor{lightgreen}{rgb}{0.832, 0.90625, 0.828125}
\definecolor{circledarkblue}{HTML}{00356C}
\definecolor{circledarkgreen}{HTML}{00B900}
\definecolor{darkblue}{HTML}{003A77}
\definecolor{darkgreen}{HTML}{008A00}
\definecolor{darkyellow}{HTML}{ef875d}
\definecolor{darkred}{HTML}{ff4033}
\definecolor{darkergreen}{HTML}{007500}
\newcommand{\rcross}[0]{\textcolor{darkred}{\faTimesCircle}}
\newcommand{\gcheck}[0]{\textcolor{darkergreen}{\faCheckCircle}}

\title[Benchmarking LLM Serving Systems for Agentic AI Workloads with \pname{}]{Benchmarking LLM Serving Systems \\for Agentic AI Workloads with \pname{}}
\begin{document}
\author{Michael Wang}
\authornote{Both authors contributed equally to this work.}
\affiliation{%
  \institution{University of Illinois}
  \city{Urbana-Champaign}
  \country{USA}
}
\author{Yikang Yue}
\authornotemark[1]
\affiliation{%
  \institution{University of Illinois}
  \city{Urbana-Champaign}
  \country{USA}
}
\author{Shaobo Li}
\affiliation{%
  \institution{University of Illinois}
  \city{Urbana-Champaign}
  \country{USA}
}
\author{Yirui Eric Zhou}
\affiliation{%
  \institution{University of Illinois}
  \city{Urbana-Champaign}
  \country{USA}
}
\author{Chen Wang}
\affiliation{%
  \institution{IBM Research}
  \country{USA}
}
\author{Jian Huang}
\affiliation{%
  \institution{University of Illinois}
  \city{Urbana-Champaign}
  \country{USA}
}

\begin{abstract}



We present \pname{}, a benchmarking framework that load-tests LLM serving systems with diverse agentic AI workloads.
It provides detailed profiling of the serving system and hardware, enabling users to identify performance bottlenecks introduced by agentic workloads.
Benchmarking LLM serving systems under agentic workloads is challenging -- 
agentic applications rely on nondeterministic LLM outputs to guide their control flow; therefore, workload patterns vary unpredictably from run to run.
\pname{} minimizes this workload variation with a fine-grained trace replay approach:
it enables users to easily collect traces from real agentic applications, synthesize new workloads with various patterns if needed, and reproducibly replay them on different LLM serving systems.
\pname{} includes eight agentic applications across diverse use cases (e.g., coding, deep research, and Q\&A) by default.
Our empirical study using these workloads shows that \pname{} accurately replays agentic workloads, provides detailed performance breakdowns, scales to larger serving systems, and assists in serving system debugging. We will open-source \pname{} on GitHub. 



\end{abstract}

\maketitle 


\section{Introduction}
With the advancement in large language models (LLMs), agentic AI applications are becoming pervasive. LLM agents can be deployed to make decisions and take actions to solve complex tasks~\cite{liang2025llmpoweredaiagentsystems}.
Solving a task can usually require more than a single LLM call: an agent issues many LLM calls to
plan~\cite{zhou2023planning}, reason~\cite{chain_of_thought}, reflect~\cite{yu2025reflect}, act on external environments via tools~\cite{react}, and coordinate with other agents~\cite{wu2024autogen}.
Their superior capabilities have driven their widespread deployment
across a growing number of use cases~\cite{pan2026measuringagentsproduction}, including coding~\cite{anthropic_claudecode_agenticcoding}, enhanced web search~\cite{odr_opendeepresearch}, data center management~\cite{erickson2024optimizing}, and scientific research~\cite{ghareeb2026multiagentresearch,schmidgall2025agentrxiv}.

As agentic applications are increasingly deployed, understanding the performance of their backend LLM serving systems becomes essential.
To rigorously evaluate LLM serving system performance, a benchmark should support diverse agentic AI workloads and enable detailed profiling of system metrics to help users identify performance bottlenecks. 


Unfortunately, existing frameworks fail to achieve these goals (see Table~\ref{tab:benchmarks}).
Agent evaluation harnesses~\cite{harbor2025,kapoor2025hal} focus on evaluating how well different agents complete tasks, rather than how the underlying serving system performs.
Although they include diverse agentic applications, they do not offer configurable load generation and cannot report system performance metrics such as latency, throughput, or hardware utilization.
Most LLM serving system performance benchmarks~\cite{llmperf_llm_inference_benchmark,reddi2020mlperf,aiperf,semianalysis_inferencex_2026} issue independent LLM calls or multi-turn conversations to the serving system, they cannot represent the typical workload patterns of agentic applications.
AA-AgentPerf~\cite{artificialanalysis_agentperf_2026}, a recent step toward benchmarking serving systems with agentic workloads, uses only proprietary coding-agent traces.
In addition, neither its benchmark nor traces are publicly available.

Furthermore, it is not easy to study the serving-system performance under agentic workloads, due to the lack of reproducibility in existing benchmarks.
Specifically, even for the same user request, the execution graph (i.e., the sequence of LLM calls, tool calls, and their dependencies) that an agentic application produces can differ drastically across runs, presenting the serving engine with a different number of LLM calls, each with different input and output token counts.

The variance occurs because agents decide their next action (e.g., which tool to call or whether to stop) based on the nondeterministic decoding output of LLMs.
Small deviations in these outputs accumulate over execution, ultimately producing different execution graphs.
We demonstrate this by running mini-SWE-agent~\cite{miniswe} with gpt-oss-120b~\cite{openai2025gptoss} twice on identical user requests under a fixed random seed: on average, the number of LLM calls produced by the same request varies by 70\% between two runs (Figure~\ref{fig:background-reproducibility}).
Therefore, to obtain reliable performance evaluations, system developers need a benchmarking framework that can measure serving system performance using reproducible agentic workloads.

 

%

To this end, we develop \pname{}, an open-source, extensible benchmarking framework that can provide detailed profiling of serving system and GPU performance metrics under various agentic workloads.
\pname{} enables users to: 
\textbf{(1)} collect workload traces from diverse agentic applications via a generic tracing module; \textbf{(2)} construct their execution graphs; 
\textbf{(3)} synthesize new workloads from collected traces to model different serving scenarios; 
\textbf{(4)} reproducibly replay these workloads against a target serving system;
and \textbf{(5)} capture detailed serving system and GPU metrics.
Specifically, \pname{} offers the following features.

\uline{First, \mbox{\pname{}} supports application-agnostic tracing of diver- se agentic workloads (\S\ref{sec:design-tracing}) to accurately construct the execu- tion graphs 
(\S\ref{sec:design-reconstruction}).}
It automatically instruments and collects traces from agentic applications developed with popular frameworks such as LangChain~\cite{langchain_sequentialchain} and LangGraph~\cite{langgraph2024} without code modifications, and requires only single-line annotations for agentic applications built from scratch.
With the traces, \pname{} efficiently constructs execution graphs using a time-span-tree-based analysis to capture the dependencies among LLM and tool calls.

\uline{Second, \pname{} can synthesize new workloads from collect- ed traces to model diverse serving
scenarios (\S\ref{sec:design-synthesis}).}
It can synthesize diverse serving system workloads such as online serving, offline batch inference, and fixed-concurrency workloads by varying the request arrival pattern over the constructed execution graphs. 
This enables users to benchmark a target LLM serving system under a wide range of realistic agentic workloads.

\uline{Third, \mbox{\pname{}} can reproducibly replay traced agentic work- loads to benchmark a target serving
system (\S\ref{sec:design-replayer}).}
For each LLM call, \pname{} submits the exact recorded input token IDs to preserve prefix-cache hit behavior, and can force the serving engine to decode the recorded output tokens, for ensuring the replay accuracy.
It issues these calls in the dependency order specified by each execution graph, taking the tool call duration into consideration,
such that the replayed workload remains realistic even when the serving engine under test differs from the one used during tracing.
This enables serving system developers to fairly and reproducibly compare different serving systems and optimizations.

 \uline{Finally, \mbox{\pname{}} can profile the serving system and under- lying hardware to obtain detailed metrics (\S\ref{sec:design-profiler}).}
It reports key serving-system metrics, including KV cache usage, prefix cache hit rate, and token throughput. 
At the hardware level, it exposes GPU metrics such as tensor core and memory bandwidth utilization. 
Together, these metrics provide a solid foundation for understanding how serving systems behave under agentic workloads. 
\pname{} collects them with a lightweight profiler which adds negligible overhead.


To demonstrate the capabilities of \pname{}, we use eight popular agentic applications that include deep research, coding, Q\&A, and general-purpose assistance (Table~\ref{tab:workloads}). They serve as ready-to-use workloads in \pname{}.
\pname{} faithfully replays agentic workloads, with a mean absolute percentage error in end-to-end request latency below 3\% for both isolated and concurrent requests (\S\ref{sec:eval-replay-valid}).

With our accurate replay, we study the end-to-end performance of agentic applications, showing that \pname{} can break down end-to-end latency (\S\ref{sec:eval-e2e-perf}) while providing detailed serving-system and hardware metrics (\S\ref{sec:eval-sys-profile}).
With this detailed profiling, \pname{} surfaces inefficiencies such as prompt designs that fail to exploit prefix caching. It helps users diagnose, from both software and hardware perspectives, serving-system performance issues such as KV cache thrashing---where concurrent requests repeatedly evict one another's cached prefixes, incurring frequent recomputation.
We further demonstrate that \pname{} can model diverse serving scenarios (\S\ref{sec:eval-serving-modes}) and scale to benchmarking multi-engine serving systems, quantifying how LLM call routing policies affect scale-out throughput (\S\ref{sec:eval-scalability}).
Additionally, \pname{} assists in serving-system debugging: by minimizing agentic workload variation, it makes the effects of even small serving-system changes clearly visible (\S\ref{sec:eval-scheduling}).
Finally, we show that \pname{} incurs minimal overhead during both trace collection and replay (\S\ref{sec:eval-overhead}).
We will open-source \pname{} for public use.

\section{Background and Motivation}
\label{sec:background-motivation}

\subsection{Agentic AI Applications}
\label{sec:applications}
At the core of agentic applications are LLM agents, which are autonomous programs that leverage LLMs to drive their decision-making and execution.
Agents can decompose and plan complex tasks~\cite{zhou2023planning}, reflect on prior experience~\cite{yu2025reflect}, collaborate with one another~\cite{wu2024autogen}, and explore external environments using tools~\cite{react}.
These capabilities power applications such as personal coding agents~\cite{anthropic_claudecode_agenticcoding}, autonomous data center managers~\cite{erickson2024optimizing}, and scientific research agents~\cite{schmidgall2025agentrxiv, ghareeb2026multiagentresearch}.

\begin{figure}
    \centering
    \includegraphics[width=0.9\linewidth]{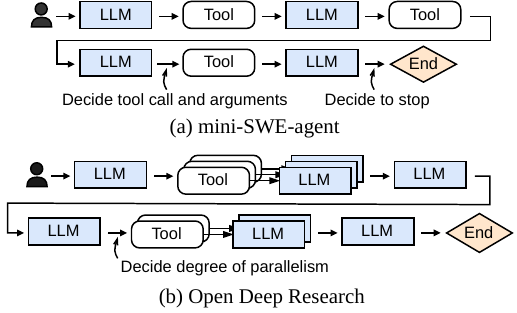}
    \vspace{-2.5ex}
    \caption{Example execution graphs of mini-SWE-agent and Open Deep Research. The graph structure is influenced by the decoding output of LLM calls.
    }
    \label{fig:agentic-characteristics}
    \vspace{-2ex}
\end{figure}

Compared to conventional chatbots, agentic applications exhibit two major differences.
First, a user request to an agentic application spawns an \emph{execution graph} of multiple dependent LLM and tool calls, rather than a single LLM call.
For example, as shown in \Cref{fig:agentic-characteristics}, mini-SWE-agent~\cite{miniswe} issues chains of alternating LLM and tool calls (i.e., ReAct~\cite{react} iterations), while Open Deep Research~\cite{odr_opendeepresearch} may produce many parallel LLM calls to summarize tool call results.
Second, agentic applications are significantly more dynamic, because their execution graphs are influenced by the nondeterministic decoding output of LLM calls, which can determine which tool to call, which agent to invoke next, and whether to stop execution.
Thus, the execution graph structure (e.g., the number of ReAct iterations and the number of parallel LLM calls) can vary significantly even for identical user requests (see detailed discussion in \S\ref{sec:challenges}). 

\subsection{System Infrastructure for Agentic Applications}
\label{sec:infrastructure}

\begin{figure}[t]
    \centering
    \includegraphics[trim=0 0 0 0.1ex,clip,width=0.85\linewidth]{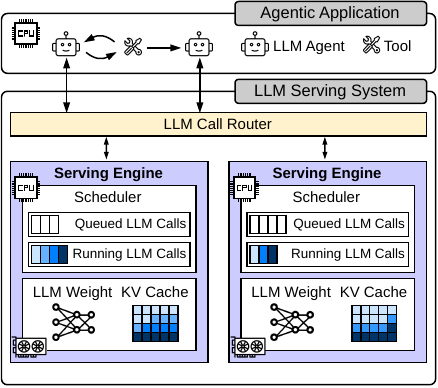}
    \vspace{-2ex}
    \caption{System infrastructure for agentic applications.}
    \label{fig:background-agent-infra}
\end{figure}

From a system perspective, agentic applications execute atop a well-abstracted LLM serving layer, as shown in ~\Cref{fig:background-agent-infra}.
    
\noindent\textbf{Agentic Application Layer.}
An agentic application is a program that defines tools, LLM agents, and a workflow that controls how they interact.
Its logic runs primarily on CPUs, with LLM calls dispatched to a separate serving system.
Each tool is a function call 
(e.g., querying a database, calling a web API, or executing code) and is exposed to an LLM through a dedicated prompt that lists each tool's purpose and the arguments it accepts. Each agent is a piece of code that assembles such prompts for an LLM call interface (e.g., the OpenAI API ~\cite{openai2024api}): it sets a system prompt defining the agent's role, includes the descriptions of the tools, and prior context such as earlier tool outputs.
The workflow, implemented either from scratch or on agentic frameworks~\cite{langchain_sequentialchain, langgraph2024, wu2024autogen}, forms a control flow graph over agents and tools: it specifies which agent/tool to execute next and how prior outputs are passed to later ones.
 As the agent-specific complexity (i.e., prompt construction and output parsing) resides at the application layer, LLM serving system sees only standard LLM calls.

\noindent\textbf{LLM Serving Layer.} 
An LLM serving system contains one or more serving engines, each of which holds the model weights and KV cache.
Serving engines efficiently execute LLM calls on the underlying hardware like GPUs. 
When an engine receives LLM calls but lacks sufficient KV cache entries to run them all, its scheduler queues the extra calls and admits them as long as enough KV cache entries become available.
Systems with multiple serving engines employ a router (e.g., vLLM-Router~\cite{vllmrouter2025}) to determine which engine should serve an LLM call based on factors such as each serving engine's load and prefix cache locality~\cite{srivatsa2025preble, vllmrouter2025, sglangrouter2024, llmd2025, dynamo2025}.

\subsection{Benchmarking LLM Serving for Agentic AI}
\label{sec:challenges}

\begingroup
\renewcommand{\arraystretch}{0.85}
\setlength{\extrarowheight}{2pt}
\begin{table*}[t!]
    \centering
    \footnotesize
    \caption{Comparison with existing benchmarks for LLM agents and LLM serving systems.
    }
    \begin{tabular}{|c|c|c|c|c|c c c|}
    \hline
    \multirow{2}{*}{\textbf{Category}} & \multirow{2}{*}{\textbf{Benchmark}} & \textbf{Diverse Agentic} & \textbf{Configurable} & \textbf{System} & \multicolumn{3}{c|}{\textbf{Reproducibility}} \\
    \cline{6-8}
     & & \textbf{Applications} & \textbf{Load Generator} & \textbf{Metrics} & \textbf{Exec. Graph} & \textbf{Input Tokens} & \textbf{Output Tokens} \\
    \hline
    \multirow{2}{*}{\vspace{-0.5ex}\shortstack{Agent\\Evaluation}}        & Harbor~\cite{harbor2025}                                  & \gcheck & \rcross & \rcross & \rcross & \rcross & \rcross \\
                                                                          & HAL~\cite{kapoor2025hal}                                  & \gcheck & \rcross & \rcross & \rcross & \rcross & \rcross \\
    \hline
    \multirow{6}{*}{\vspace{-0.5ex}\shortstack{Serving System\\Performance}} & InferenceX~\cite{semianalysis_inferencex_2026}            & \rcross & \gcheck & \gcheck & \rcross & \gcheck & \rcross \\
                                                                          & LLMPerf~\cite{llmperf_llm_inference_benchmark}            & \rcross & \gcheck & \gcheck & \rcross & \gcheck & \rcross \\
                                                                          & MLPerf~\cite{reddi2020mlperf}                             & \rcross & \gcheck & \gcheck & \rcross & \gcheck & \rcross \\
                                                                          & AIPerf~\cite{aiperf}                                      & \rcross & \gcheck & \gcheck & \rcross & \gcheck & \rcross \\
                                                                          & AA-AgentPerf~\cite{artificialanalysis_agentperf_2026}     & \rcross & \rcross & \gcheck & \gcheck & \gcheck & \rcross \\
                                                                          & \textbf{\pname{} (Ours)}                                  & \gcheck & \gcheck & \gcheck & \gcheck & \gcheck & \gcheck \\
    \hline
    \end{tabular}
\label{tab:benchmarks}
\end{table*}
\endgroup



Today’s mainstream serving systems, such as vLLM~\cite{vllm}, are designed primarily for independent LLM calls.
Consequently, they suffer from suboptimal LLM call scheduling~\cite{agentix, kairos} and KV cache management~\cite{agserve2025, li2025continuum} when serving agentic workloads.
To understand and mitigate these inefficiencies, it is essential to benchmark serving systems under realistic agentic workloads, allowing developers to accurately identify system bottlenecks and
test proposed optimizations.

\begin{figure}[t]
    \begin{subfigure}[b]{0.18\textwidth}
        \centering
        \includegraphics[width=\linewidth]{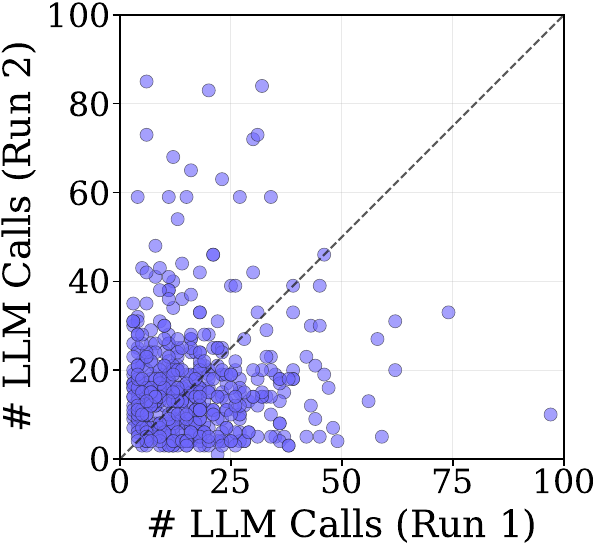}
        \caption{LLM calls per request. Each dot represents two runs of the same request.
        }
        \label{fig:llm_scatter-no-replay}
    \end{subfigure}
    \hfill
    \begin{subfigure}[b]{0.27\textwidth}
        \centering
        \includegraphics[width=\linewidth]{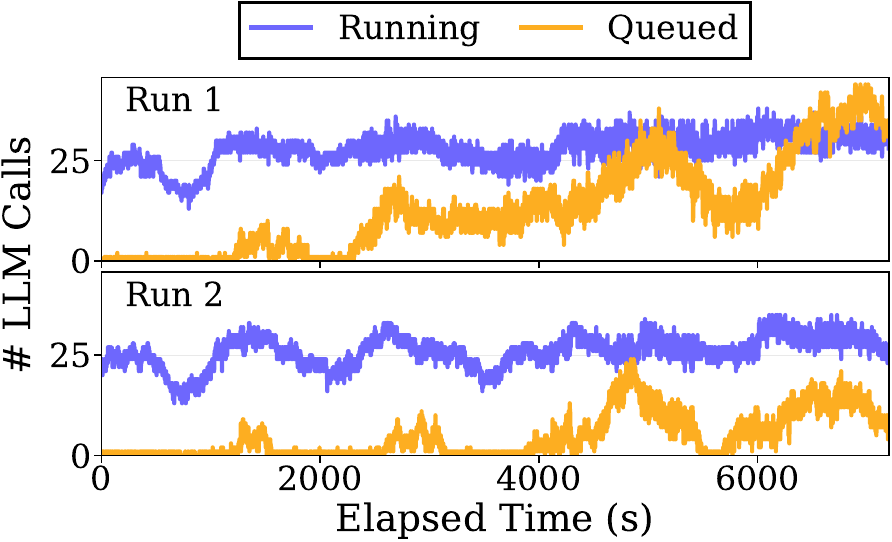}
        \caption{Running and queued LLM calls in the serving system across two runs over the same two-hour time window.}
        \label{fig:request-timeseries-no-replay}
    \end{subfigure}
    \vspace{-2
    ex}
    \caption{
    Run-to-run variance of agentic applications, illustrated with mini-SWE-agent. Results are profiled on SWE-bench, served by gpt-oss-120b online at 4.5~requests/minute.
    }
    \label{fig:background-reproducibility}
\end{figure}

Conducting such 
benchmarks requires reproducibly stress-testing the serving system with diverse agentic applications. 
However, the agentic applications have dynamic execution graphs, making reproducible benchmarking fundamentally challenging.
Specifically, even when identical user requests are executed with a fixed random seed,
the resulting execution graphs differ substantially across runs.

For example, as shown in Figure~\ref{fig:llm_scatter-no-replay}, when running mini-SWE-agent~\cite{miniswe} on SWE-bench~\cite{swebench}, the total number of LLM calls triggered by identical user requests varied by an average of 12.9 calls between two runs (76\% of the average number of LLM calls per request).
Thus, even with identical user requests, arrival times, and random seeds\footnote{Even with a fixed random seed, decoding output may diverge across executions because batched inference can introduce minute floating-point differences~\cite{yuan2026understanding}. 
Although employing batch-invariant kernels can remove this source of nondeterminism, it typically incurs substantial overhead~\cite{he2025nondeterminism}.}, accumulated request-level variance significantly alters the actual workload during benchmarking.
\Cref{fig:request-timeseries-no-replay} shows the overall trend in queued LLM calls diverging significantly across two runs within a two-hour window after warm-up.
This divergence yields significantly different benchmarking outcomes.
Within the two-hour observation window, run 2 exhibits 65\% less cumulative queuing delay across all LLM calls than run 1. 
Correspondingly, run 2's average end-to-end request latency is 25\% lower, while the 95th percentile (P95) latency is 30\% lower. 
It is hard to isolate the effect of system optimizations from the variance of workload patterns. 

Unfortunately, no existing benchmark addresses the above challenge (see~\Cref{tab:benchmarks}).
Agent evaluation benchmarks~\cite{kapoor2025hal, harbor2025} focus on evaluating the ability of agents to complete tasks rather than the performance of the underlying serving system.
They include a variety of agentic applications, but offer no configurable load generator for stress-testing the serving system, expose no system-level performance metrics,
and cannot reproduce a consistent workload across runs.

LLM serving benchmarks~\cite{llmperf_llm_inference_benchmark, reddi2020mlperf, aiperf,semianalysis_inferencex_2026} offer users with configurable load generators and system profiling, but their input prompts are sampled from datasets or synthesized from user-specified distributions.
Such independently drawn prompts cannot replicate the unique behavior of agentic workloads -- a single request expands into many interdependent LLM calls whose prompts are assembled from earlier outputs and tool results.
AA-AgentPerf~\cite{artificialanalysis_agentperf_2026} worked toward performance benchmarking for agentic serving systems, but it is a closed ranking service.
Developers can only submit a system to the service provider and receive a published ranking, it prevents developers from reproducing results, adding new applications, and identifying system bottlenecks. Moreover, its fixed workloads may not reflect what developers deploy in their own infrastructure.

These limitations motivate us to build an open-source and extensible benchmarking framework for agentic serving.

\section{Design and Implementation}
%




\begin{figure}[t]
    \includegraphics[width=0.8\linewidth]{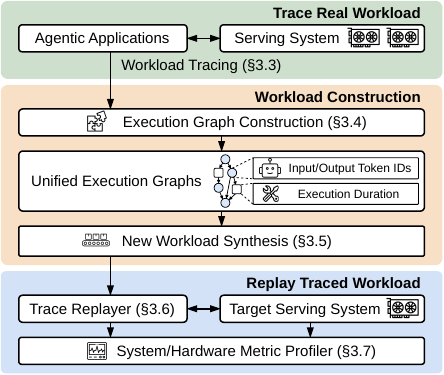}
    \caption{System overview of \pname{}.}
    \label{fig:system-overview}
\end{figure}

\subsection{Goals of \pname{}}
We develop \pname{} with the following goals:
\begin{itemize}[leftmargin=*,topsep=2pt,partopsep=-1.2pt]
    \item \textbf{Reproducible execution.} \pname{} should benchmark serving systems with reproducible agentic workloads, isolating serving system performance from the run-to-run variance of agentic applications.
    \item \textbf{Workload diversity.} \pname{} should support a wide range of agentic applications and diverse serving scenarios such as online serving and offline batch inference.
    \item \textbf{System metric collection.} \pname{} should provide a set of lightweight profiling tools to collect detailed system metrics across the full stack, from the agentic application to the serving system to the underlying hardware, such as end-to-end user request latency, prefix cache hit rate, GPU hardware utilization, and others.
    \item \textbf{Portability.} \pname{} should be compatible with agent development frameworks, allowing users to easily extend it with new applications.
    It should also be able to evaluate diverse serving systems, as well as larger-scale deployments with multiple serving engines.
    \item \textbf{Ease of Use.} 
    \pname{} should minimize developer burden throughout the end-to-end benchmarking workflow, from integrating a new agentic application to generating new workload patterns and profiling system performance.
\end{itemize}


\subsection{System Overview}


\pname{} achieves these goals via a trace replay approach, as shown in Figure~\ref{fig:system-overview}.
To collect agentic workload traces, a user first instruments the target application with \pname{}. The user then runs the application while \pname{} traces its LLM calls and tool calls (\S\ref{sec:design-tracing}).
Based on the trace, \pname{} first constructs the execution graph for each user request (\S\ref{sec:design-reconstruction}).
It then generates new workload patterns (if needed) by deciding when to replay each graph, either following user-specified request arrival patterns or replaying the exact arrival times observed during tracing (\S\ref{sec:design-synthesis}).
During trace replay, \pname{} load-tests the target LLM serving system by replaying these execution graphs. 
It traverses each graph to issue LLM calls in dependency order, such that the replayed graph reproduces the behavior of the original request (\S\ref{sec:design-replayer}).
Meanwhile, the system profiler (\S\ref{sec:design-profiler}) monitors execution and collects system performance metrics.




\subsection{Workload Tracing}
\label{sec:design-tracing}

\begin{figure}[t]
    \includegraphics[trim=0 0 0 1.1ex,clip,width=0.87\linewidth]{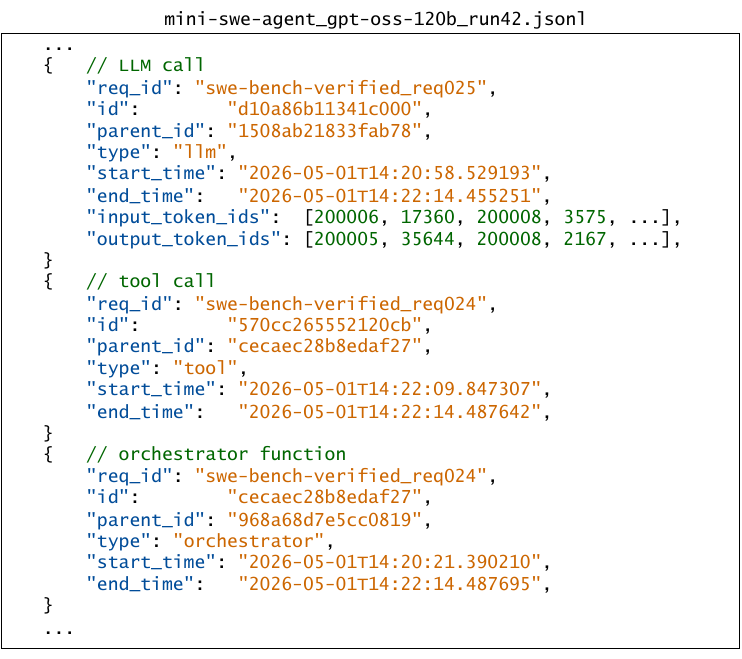}
    \vspace{-3ex}
    \caption{Example trace collected by \pname{}.}
    \label{fig:raw-trace}
\end{figure}

\pname{} employs a tracing submodule to collect the information needed for workload replay.
However, real-world agentic applications are developed with different frameworks or even built from scratch, they lack a common interface to collect application traces.
The \pname{} tracing module generalizes across diverse applications while requiring minimal changes to their code.
It decomposes the agentic application's execution into two operation types, LLM calls and tool calls, and traces each separately:

\noindent\textbf{LLM calls.}
Agentic applications typically make LLM calls through standardized interfaces such as OpenAI APIs~\cite{openai2024api} and Anthropic SDKs~\cite{anthropic-python-sdk}.
\pname{} traces all LLM calls issued through these interfaces via library interposition~\cite{interposition}.
Specifically, \pname{} wraps the LLM call interface in the shared client library, so every LLM call will first enter the \pname{} wrapper function.
Once the call returns, \pname{} records the input and output token IDs the serving engine returns for each LLM call.
It also employs OpenTelemetry~\cite{opentelemetry} to record the start and end time of each call, as well as the user request it belongs to.
Because the interposition occurs within shared client libraries, it requires no changes to the application code.



\noindent\textbf{Tool calls.}
In contrast to LLM call interfaces, tool calls lack a standardized interface.
A tool call can be a bash command~\cite{miniswe}, a web search API~\cite{odr_opendeepresearch}, or a custom code segment through which an agent makes a decision~\cite{zhou2023planning}.
To support tracing different kinds of tool calls, 
for tool calls defined under popular agentic frameworks such as LangChain~\cite{langchain_sequentialchain} and LangGraph~\cite{langgraph2024}, \pname{} uses the same library interposition technique, wrapping the tool calls with \pname{} tracing functions without any application changes.
For applications developed from scratch in Python, \pname{} provides a single-line annotation API that allows users to manually wrap each tool call.
With these annotations, \pname{} likewise traces the start and end time of each tool call as well as the user request it belongs to via OpenTelemetry.
As a safeguard against missed annotations, whenever a new LLM or tool call is made, \pname{} also checks whether any other call completed within the preceding window (20\,ms by default).
If none did, \pname{} reports the gap as a likely unannotated region. This increases observability of the application by highlighting unnoticed overhead.

In addition to LLM and tool calls, we also introduce \textit{orchestrator functions} to trace how these calls are nested, facilitating more accurate execution graph construction.
We define orchestrator functions as functions that compose LLM calls, tool calls, or smaller orchestrator functions into more complex logic.
They are common in applications built with agentic frameworks.
For example, LangChain~\cite{langchain_sequentialchain} and LangGraph~\cite{langgraph2024} wrap each LLM call and tool call in a \texttt{Runnable} and compose these into a chain or graph. 
These chains or graphs itself is also a \texttt{Runnable} class.
Because these frameworks expose all orchestrator functions through a common class such as \texttt{Runnable}, \pname{} interposes on that class to trace the start and end time of every orchestrator function once it detects one of these frameworks.


\pname{} also records the call hierarchy. For each LLM call, tool call, and orchestrator function, \pname{} uses OpenTelemetry to record the parent orchestrator function that encloses it.
For each agentic application run, \pname{} logs all information to a JSONL file, as illustrated in \Cref{fig:raw-trace}.
The trace records the start and end time and the parent ID of every user request, LLM call, tool call, and orchestrator function. For LLM calls, we also record the input and output token IDs.
This tracing incurs negligible latency overhead, which we quantify in \S\ref{sec:eval-overhead}.



\begin{figure}[t]
    \begin{subfigure}[b]{\linewidth}
        \centering
        \includegraphics[trim=0 0 0 0.2ex, clip, width=\linewidth]{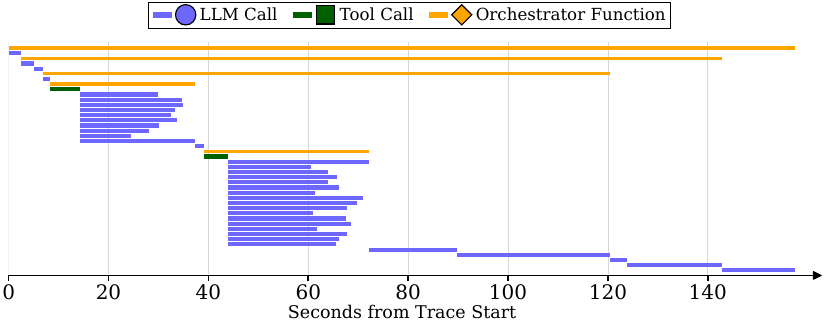}
        \caption{Timeseries of spans.}
        \label{fig:design-timeseries-span}
    \end{subfigure}
    \begin{subfigure}[b]{0.49\linewidth}
        \centering
        \includegraphics[width=\linewidth]{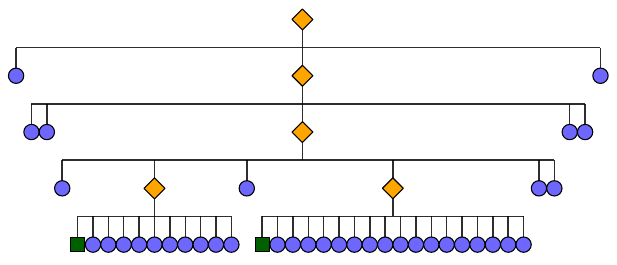}
        \caption{Time-span tree.}
        \label{fig:design-tree}
    \end{subfigure}
    \hfill
    \begin{subfigure}[b]{0.49\linewidth}
        \centering
        \includegraphics[width=\linewidth]{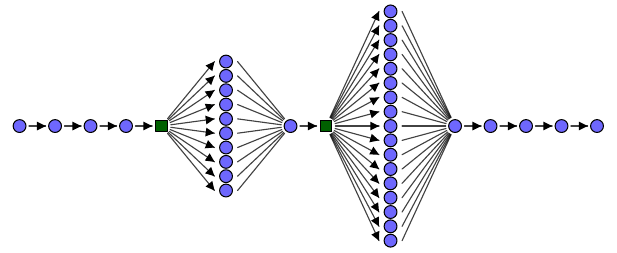}
        \caption{Execution graph.}
        \label{fig:design-dag}
    \end{subfigure}
    \vspace{-2ex}
    \caption{Execution graph construction.}
    \label{fig:design-span-tree-dag}
\end{figure}

\subsection{Execution Graph Construction}
\label{sec:design-reconstruction}

\newlength{\commentcol}
\setlength{\commentcol}{5cm}   
\newcommand{\lc}[2]{%
  \DontPrintSemicolon%
  \makebox[\commentcol][l]{$#1$\,;}\tcp*[h]{#2}\;%
  \PrintSemicolon%
}

\begin{algorithm}[t]
\footnotesize
\linespread{0.88}\selectfont
\caption{Execution graph construction}
\label{alg:reconstruct}
\SetArgSty{textnormal}
\SetCommentSty{algcommentfont}
\SetKwInOut{Input}{Input}
\SetKwInOut{Output}{Output}
\SetKwFunction{Build}{BuildGraph}
\SetKwFunction{DepSpans}{DependentSpans}
\SetKwFunction{InterEdges}{InterGraphEdges}
\SetKwFunction{Reduce}{TransitiveReduction}
\SetKwProg{Fn}{Function}{:}{end}
\Input{Root span $s$ of a time-span tree; each span $c$ exposes start time $c.\mathit{start}$, end time $c.\mathit{end}$, and children $c.\mathit{children}$}
\Output{Execution graph $G=(V,E)$}
\Fn{\DepSpans{$S$}}{
  $P \gets \{\,(s_1,s_2) \in S \times S : s_1.\mathit{end} \le s_2.\mathit{start}\,\}$\;
  \tcp{drop $(a,b)$ when $a$ reaches $b$ transitively}
  \Return \Reduce{$P$}\;
}
\Fn{\InterEdges{$G_1,\,G_2$}}{
  \lc{T \gets \{\, u \in G_1.V : \deg^{+}_{G_1}(u) = 0 \,\}}{sinks of $G_1$}
  \lc{R \gets \{\, v \in G_2.V : \deg^{-}_{G_2}(v) = 0 \,\}}{sources of $G_2$}
  \Return $\{\,(u,v) : u \in T,\ v \in R\,\}$\;
}
\Fn{\Build{$s$}}{
  \lc{\KwSty{if}\ s.\mathit{children}=\emptyset\ \KwSty{then}\ \KwSty{return}\ (\{s\},\,\emptyset)}{LLM/tool call}
  $\mathcal{G} \gets \emptyset$;\quad $V \gets \emptyset$;\quad $E \gets \emptyset$;
  
  \ForEach{$c \in s.\mathit{children}$}{
    $\mathcal{G}[c] \gets{}$ \Build{$c$}\;
    $V \gets V \cup \mathcal{G}[c].V$;\quad $E \gets E \cup \mathcal{G}[c].E$\;
  }
  \ForEach{$(c_1,c_2) \in{}$ \DepSpans{$s.\mathit{children}$}}{
    $E \gets E \cup{}$ \InterEdges{$\mathcal{G}[c_1],\, \mathcal{G}[c_2]$}\;
  }
  \Return $(V,E)$\;
}
\end{algorithm}

Obtaining the traces shown in \S\ref{sec:design-tracing} itself is insufficient. If we naively replay a workload trace by issuing each LLM call with its recorded token IDs at its recorded time, it fails to reflect the dependencies among LLM calls.
Whenever the serving system or the underlying hardware differs from the setup used during tracing, the replayed workload becomes unrealistic.
For example, an improved engine that decodes tokens faster finishes each LLM call sooner, so the calls that depend on it should also be issued earlier.
To accurately reproduce the behavior of agentic applications, it is essential to construct the dependencies between the LLM and tool calls triggered by each user request.

\pname{} models the execution of each user request as a directed acyclic graph (DAG), in which each node is an LLM call or a tool call.
A directed edge $a \rightarrow b$ indicates that call $b$ depends on the result of call $a$ and thus can execute only after $a$ completes. To build the DAG of each user request, \pname{} must infer the dependencies between calls.

Fortunately, the trace already records every operation's time span with its parent span.
\pname{} links each span to its parent to build a \emph{time-span tree} that recovers the application's structure.
Figure~\ref{fig:design-timeseries-span} shows example spans traced from Open Deep Research, and Figure~\ref{fig:design-tree} shows the tree built from them.
LLM call and tool call spans form the leaf nodes, since they perform the actual work and invoke no further calls.
Their parent spans, which are orchestrator functions (\S\ref{sec:design-tracing}), form the internal nodes, and the root node is the span of the entire user request.

With the time-span tree, \pname{} further builds the execution graph of the application request.
We show the graph construction algorithm in Algorithm~\ref{alg:reconstruct}.
Specifically, for each leaf node, which is either the time span of an LLM or a tool call, it produces a single-node DAG (line~11).
For each non-leaf node, it first builds subgraphs for its children (line~13) and identifies dependencies between them (line~16).
\pname{} treats two spans as dependent when one finishes before the other starts (line~2), and keeps only direct dependencies (line~3).
For each pair of dependent spans, \pname{} connects their subgraphs by adding edges (line~17) from every call in the earlier span that has no successor (line~6) to every call in the later span that has no predecessor (line~7).
Overall, \pname{} gradually connects the subgraphs of the child subtrees into larger graphs, ultimately producing the entire execution graph for each user request.
Figure~\ref{fig:design-dag} shows the execution graph constructed from Figure~\ref{fig:design-tree}.

Using the time-span tree information is important to avoid constructing false dependencies.
For example, LATS~\cite{zhou2023planning} launches multiple independent parallel tasks, each containing one LLM call followed by a tool call.
Thus, inferring dependencies without consulting the time-span tree (i.e., applying \texttt{DependentSpans} directly to all calls) introduces false dependencies: when two LLM calls in different tasks finish at nearly the same time, the tool call in one task can mistakenly treat an LLM call in another task as its predecessor.
\pname{} avoids this issue because the calls from different tasks will be assigned as leaf nodes under different internal nodes (e.g., different orchestrator functions) in the time-span tree.
It can correctly construct the execution graph of all applications used in our evaluation.

\pname{} efficiently constructs DAGs for diverse agentic applications.
The traces we collected in our evaluation contain no more than 1000 LLM/tool calls per user request.
Each DAG can be constructed in less than 200\,ms using a single thread on an AMD EPYC 9334 32-core CPU.

\subsection{New Workload Synthesis}
\label{sec:design-synthesis}
By default, \pname{} replays each request at the exact arrival time observed during tracing.
However, users often need to benchmark a serving system under serving scenarios beyond the traced one, such as online serving at a higher request rate or offline batch inference.
Rather than collecting a new trace for each scenario, \pname{} allows the synthesis of new workloads by varying the request arrival pattern across the same set of DAGs.
  
First, \pname{} can generate online serving workloads at a target request rate by sampling arrival times from a Poisson process.
Second, \pname{} can generate an offline batch inference~\cite{sheng2023flexgen} workload by dispatching a specified number of requests at the start of execution.
Third, \pname{} supports maintaining a fixed number of concurrent requests (i.e., fixed-concurrency serving) to synthesize a workload in which multiple users share one serving system~\cite{artificialanalysis_agentperf_2026}.
Finally, \pname{} supports custom request arrival patterns, specified by a user-provided configuration file that lists each request ID and its arrival time.

By default, \pname{} treats tool-call durations as independent of serving system load, since tool calls typically execute on a separate machine.
However, \pname{} lets users configure tool call durations to study their impact on serving system behavior and end-to-end request performance.

\subsection{Replaying Traces}
\label{sec:design-replayer}

We now describe how \pname{} replays the traced workload bottom-up: first, how \pname{} replays individual LLM calls, then how it replays each execution graph, and finally, how it replays the entire workload.

\noindent\textbf{Replaying an LLM call.}
With the collected input/output token IDs, \pname{} supports two LLM call replay modes.
In the first mode, \pname{} sends the recorded input token IDs to the serving engine to preserve the prefix-cache-hitting behavior and instructs it to ignore the \texttt{EOS} token (e.g., set \texttt{ignore\_eos} for vLLM).
The engine therefore decodes until it reaches \texttt{max\_tokens}, which \pname{} sets to the recorded output sequence length.
This ensures the engine decodes the same number of tokens recorded in the trace.
The second mode further forces the serving engine to generate exactly the recorded output token sequence.
This benefits benchmarking when serving LLMs with mixture-of-experts~\cite{lepikhin2021gshard} or native sparse attention~\cite{yuan2025native}, as the generated token IDs influence expert routing and attention sparsity.
The first mode does not require any modification to the serving engine, while the second requires minor changes. We detail our implementation of output token ID enforcement in \S\ref{sec:impl}.

\noindent\textbf{Replaying an execution graph.}
Execution graph replay follows DAG semantics: a node executes as soon as all its predecessors finish.
The trace replayer therefore launches each node as an asynchronous task that suspends until all its predecessor calls complete.
When awoken, an LLM call node replays its LLM call and waits for the call to finish, while a tool call node further sleeps for the traced tool call duration to reproduce its latency.



\noindent\textbf{Replaying the entire workload.}
\pname{} replays the entire workload by replaying each execution graph at the time specified by the workload synthesizer, running multiple in-flight execution graphs via multiprocessing.
For workloads with many requests, holding all execution graphs in memory is costly, so \pname{} loads them on the fly: a fixed-size buffer prefetches execution graphs for upcoming requests and frees memory for finished ones.
Its memory footprint therefore depends only on the number of concurrent requests, not the total workload size.


\subsection{System Profiling}
\label{sec:design-profiler}



\pname{} integrates a system profiler to collect serving-system and hardware metrics. 
For serving-system metrics, it polls the HTTP \texttt{/metrics} endpoint exposed by most popular serving systems, capturing (1) the number of running and queued LLM calls, (2) KV cache capacity and prefix cache hit rate, and (3) input and output token throughput.
For GPU metrics, the profiler uses NVIDIA's CUDA Profiling Tools Interface~\cite{cupti} (CUPTI). \
It measures utilization of
(4) tensor cores, (5) streaming multiprocessors and (6) memory bandwidth. 
The metric sampling interval is set as 1 second for the serving system and 20\,ms for the GPU and is configurable. We show the profiler has low overhead in \S\ref{sec:eval-overhead}.

\section{Implementation Details}
\label{sec:impl}


\pname{} provides eight containerized, instrumented agentic applications to reduce environment setup burden (see \S\ref{sec:eval-workloads}).
It is implemented in Python as four modules: application tracing, execution graph construction, workload synthesis, and trace replay. 
The tracing module configures the OpenTelemetry SDK~\cite{opentelemetry} with automatic instrumentation to generate spans, which are exported over OTLP~\cite{otel-otlp}/gRPC~\cite{grpc} to Jaeger~\cite{jaeger}, an OpenTelemetry-compatible tracing backend. 
The system profiler runs as a background process, periodically sampling serving-system metrics from each engine's HTTP endpoint and GPU metrics via CUPTI~\cite{cupti}. For multi-node deployments, \pname{} can deploy the profiler to each node via Ansible~\cite{ansible} to report per-GPU metrics. 



We prototype output token ID enforcement in vLLM~\cite{vllm}.
In the messages \pname{} sends to vLLM for each LLM call, \pname{} appends an extra message that carries the output token IDs and assigns it a special role \texttt{replayer}, distinguishing it from the normal \texttt{system}, \texttt{user}, and \texttt{assistant} roles.
vLLM moves this special message into the per-call sampling parameters and drops it from the message list.
After each decoding iteration, vLLM overrides the sampled tokens with the supplied token IDs.
To support speculative decoding~\cite{leviathan2023fast}, \pname{} instructs the serving engine to report the number of tokens drafted successfully in each decoding iteration during trace collection.
To replay, \pname{} sends the per-iteration accepted-token counts through the same special message, and vLLM accepts tokens accordingly.
We find no measurable overhead from output token ID enforcement.
\section{Evaluation}

Our evaluation shows that \pname{} can
(1) accurately replay the serving system workload produced by diverse agentic applications (\S\ref{sec:eval-replay-valid});
(2) collect end-to-end performance metrics for each user request and provide a detailed breakdown (\S\ref{sec:eval-e2e-perf});
(3) profile detailed serving system metrics and hardware utilization metrics (\S\ref{sec:eval-sys-profile});
(4) synthesize workloads for diverse serving scenarios (\S\ref{sec:eval-serving-modes}); and
(5) scale to multi-engine serving systems (\S\ref{sec:eval-scalability}).
We also demonstrate how \pname{}'s reproducible trace replay helps developers debug serving systems for agentic applications (\S\ref{sec:eval-scheduling}).
Finally, we show that \pname{} incurs minimal overhead during both trace collection and replay (\S\ref{sec:eval-overhead}).

\begin{table*}[hbt!]
\footnotesize
\centering
\caption{Agentic workloads used in our evaluation. LLM-call counts are per user request and profiled using gpt-oss-120b.}
\vspace{-3ex}
{ 
\setlength{\tabcolsep}{3pt}
\begin{tabular}{|c|c|c|c|c|%
>{\raggedleft\arraybackslash}m{0.6cm}|>{\raggedleft\arraybackslash}m{0.6cm}|}
\hline
\multirow{2}{*}{\textbf{Application}} &
\multirow{2}{*}{\textbf{Framework}} &
\multirow{2}{*}{\textbf{Category}} &
\multirow{2}{*}{\textbf{Tools}} &
\multirow{2}{*}{\textbf{Dataset}} &
\multicolumn{2}{c|}{\textbf{LLM Calls}} \\
\cline{6-7}
& & & & & \textbf{Avg.} & \textbf{P95} \\
\hline
\makecell{Open Deep Research~\cite{odr_opendeepresearch}} &
\makecell{LangGraph} &
\makecell{Deep Research} &
Web search API &
\makecell{ResearchyQuestions~\cite{researchyquestions}} &
69.6 & 143.8 \\
\hline
\makecell{DeerFlow~\cite{deerflow2025}} &
\makecell{LangGraph} &
\makecell{Deep Research} &
Web search API &
\makecell{ResearchyQuestions~\cite{researchyquestions}} &
14.7 & 19.0 \\
\hline
\makecell{mini-SWE-agent~\cite{miniswe}} &
\makecell{Custom} &
\makecell{Coding} &
Terminal &
\makecell{SWE-Bench~\cite{swebench}} &
17.0 & 39.0 \\
\hline
\makecell{LATS~\cite{zhou2023planning}} &
\makecell{LangGraph} &
\makecell{Q\&A} &
Web search API &
\makecell{HotpotQA~\cite{yang2018hotpotqa}} &
10.6 & 26.0 \\
\hline
\makecell{LLMCompiler~\cite{llmcompiler}} &
\makecell{LangChain} &
\makecell{Q\&A} &
Web search API, Terminal &
\makecell{ParallelQA~\cite{llmcompiler}} &
2.4 & 6.0 \\
\hline
\makecell{Tau-Bench~\cite{taubench}} &
\makecell{Custom} &
\makecell{General} &
Database &
\makecell{Airline, Retail~\cite{taubench}} &
14.9 & 26.0 \\
\hline
\makecell{CUGA~\cite{marreed2025cuga}} &
\makecell{LangGraph} &
\makecell{General} &
Web search API, Terminal &
\makecell{AssistantBench~\cite{assistantbench}} &
7.3 & 17.0 \\
\hline
\makecell{MagenticOne~\cite{magentic_one}} &
\makecell{AutoGen} &
\makecell{General} &
Web browser, Terminal &
\makecell{AssistantBench~\cite{assistantbench}} &
43.7 & 54.1 \\
\hline
\end{tabular}
} 
\label{tab:workloads}
\vspace{-2ex}
\end{table*}

\begin{figure*}[t]
    \begin{subfigure}{0.4\linewidth}
        \centering
        \includegraphics[width=\linewidth]{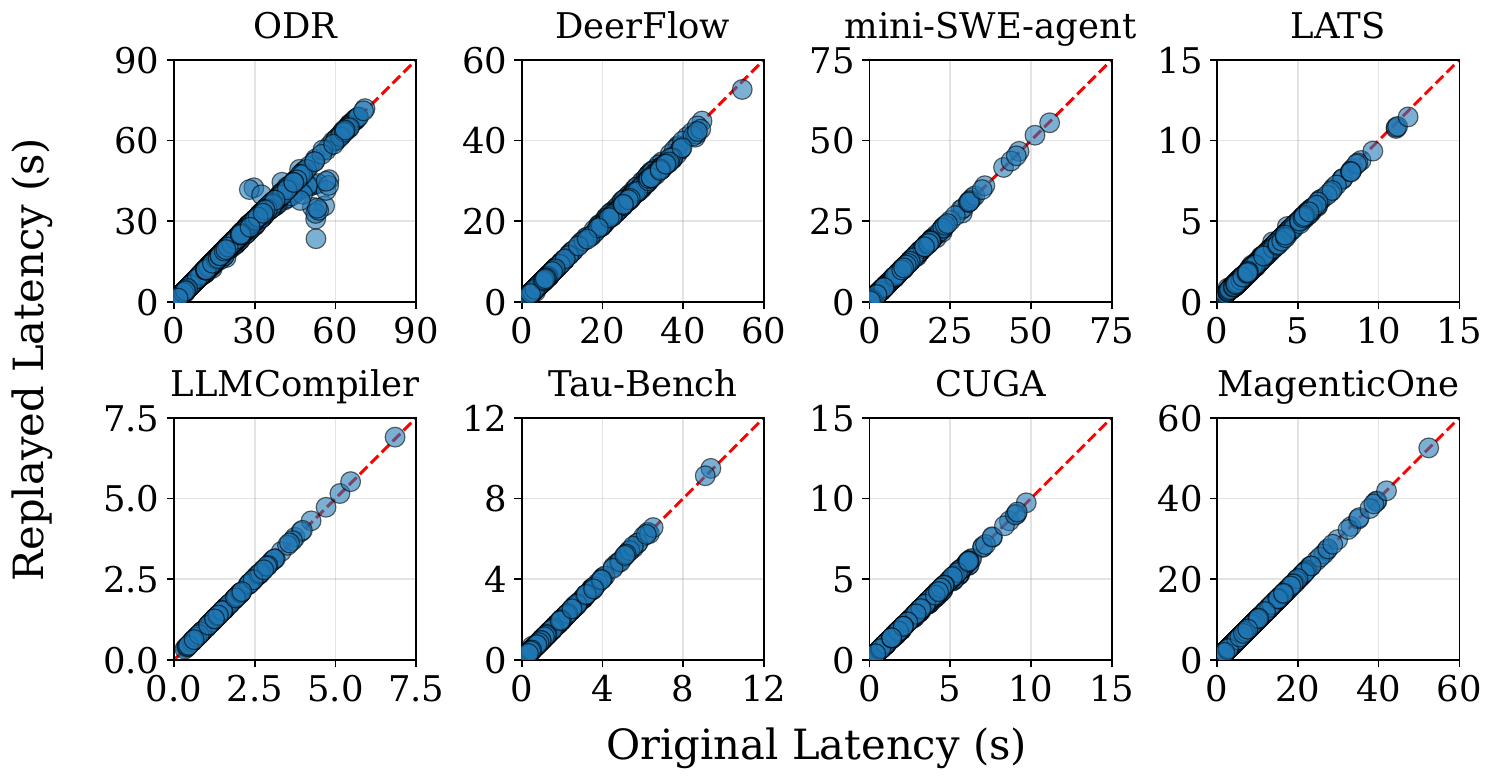}
        \vspace{-4.5ex}
        \caption{Individual LLM call latency.}
        \label{fig:replay-val-iso-llm}
    \end{subfigure}
    \hfill
    \begin{subfigure}{0.4\linewidth}
        \centering
        \includegraphics[width=\linewidth]{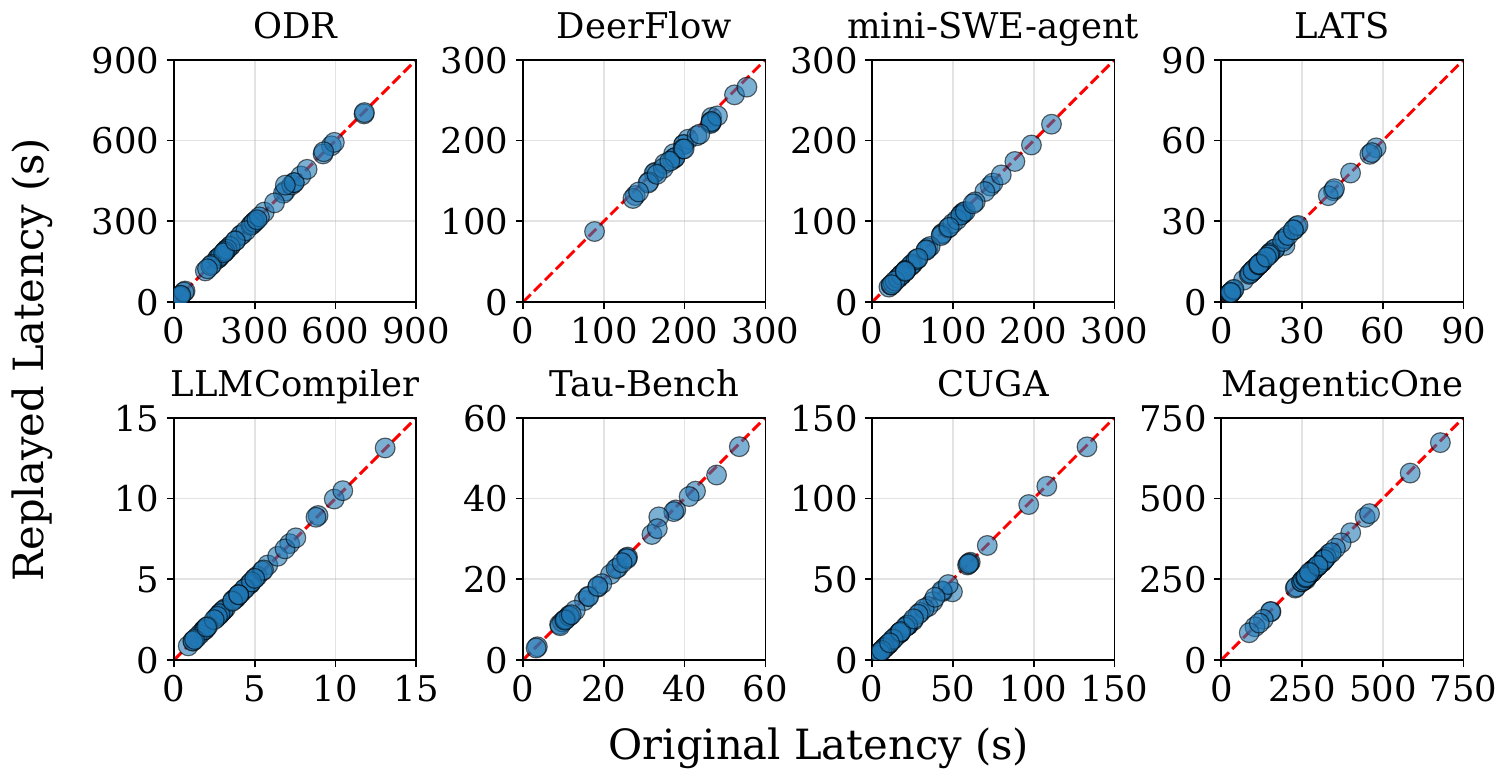}
        \vspace{-4.5ex}
        \caption{End-to-end user request latency.}
        \label{fig:replay-val-iso-request}
    \end{subfigure}
    \hfill
    \begin{subfigure}[b]{0.18\linewidth}
        \centering
        \includegraphics[width=\linewidth]{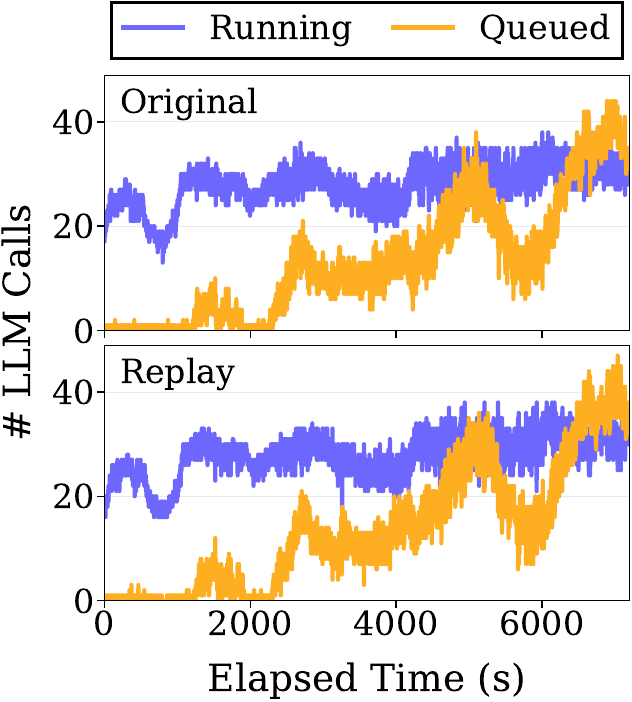}
        \vspace{-4.25ex}
        \caption{Online serving.}
        \label{fig:request-timeseries-replay}
    \end{subfigure}
    \vspace{-2.5ex}
    \caption{Validation of replay fidelity. }
    \vspace{-2ex}
\end{figure*}

\subsection{Experimental Setup}
\label{sec:eval-setup}

\noindent\textbf{LLM.} We use gpt-oss-120b~\cite{openai2025gptoss} and Qwen3.6-35B-A3B~\cite{qwen36_35b_a3b} as the LLM backbones for our agentic applications. Both are state-of-the-art mixture-of-experts models explicitly post-trained for tool usage.
By default, we use gpt-oss-120b.

\noindent
\textbf{Serving System.} We use vLLM v0.19.0~\cite{vllm} as the serving engine and vLLM Router~\cite{vllmrouter2025} to route LLM calls to different engines in multi-engine deployments.
We apply minor modifications to vLLM to enable CPU prefix caching for Qwen3.6-35B-A3B and output token ID enforcement.

\noindent \textbf{Testbed.} Our testbed is a server with dual AMD EPYC 9334 32-core CPUs with 1.5TB of DDR5 memory. The server has two NVIDIA H100 NVL 94GB GPUs, each attached to the host through a PCIe Gen5 x16 link.
We use two such servers connected by a 100Gbps InfiniBand switch for scalability experiments in \S\ref{sec:eval-scalability}.



\subsection{Agentic Applications}
\label{sec:eval-workloads}

Our evaluation uses diverse agentic workloads covering a variety of popular use cases, as shown in \Cref{tab:workloads}.

\noindent\textbf{Deep Research.} Deep research applications conduct large-scale web searches to answer complex user queries in detail.
LLM calls are used to produce search queries and summarize search results.
Open Deep Research (ODR) launches multiple web searches in parallel, each followed by an LLM call that summarizes its results; on average, 20.1 such search-and-summarize operations run concurrently.
DeerFlow instead makes sequential LLM calls, using each call to summarize multiple search results. Compared to ODR, it produces fewer LLM calls, each with significantly longer prompts. 



\noindent\textbf{Coding.}
We use mini-SWE-agent as a representative coding agent. It follows the ReAct paradigm, alternating between LLM calls and terminal commands to resolve GitHub issues from popular open-source repositories.
Its LLM calls expose substantial opportunities for prefix-cache sharing: the prompt for each LLM call concatenates the inputs and outputs of all preceding LLM and tool calls.

\noindent\textbf{Q\&A.} We select two representative Q\&A applications, LATS and LLMCompiler.
LATS issues relatively more LLM calls, as it explores and evaluates many candidate answers in parallel before settling on a final one, while LLMCompiler issues fewer: a single LLM call plans a graph of downstream calls, which are then executed without further planning.


\noindent\textbf{General.}
We also include three general-purpose applications.
Tau-Bench uses a single ReAct agent for customer service, with tools to query and update customer records in a database.
CUGA builds a generalist ReAct agent with code execution and web search tools.
MagenticOne coordinates multiple agents, raising its LLM-call count substantially.


\begin{figure}
    \includegraphics[width=\linewidth]{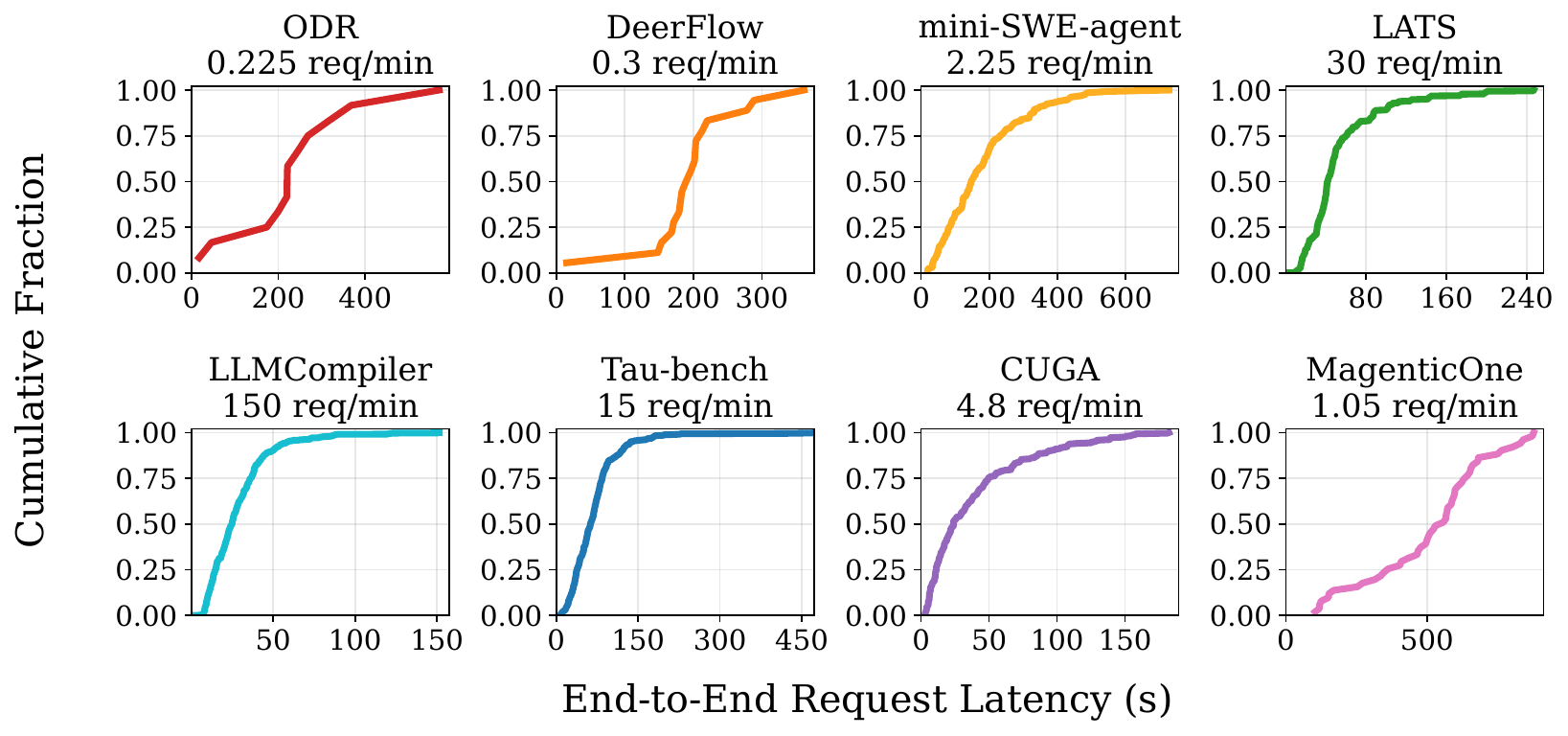}
    \vspace{-6ex}
    \caption{CDF of end-to-end latency per user request.}
    \label{fig:eval-latency-cdf}
    \vspace{-2ex}
\end{figure}

\begin{figure*}
    \hspace*{0.3cm}
    \includegraphics[trim=0 0 0 1.8ex,clip,width=\dimexpr\linewidth-0.35cm\relax]{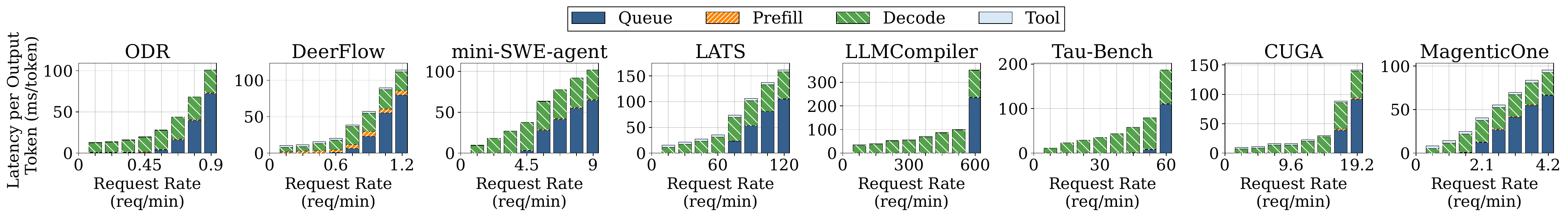}
    \vspace{-5ex}
    \caption{Latency breakdown per output token.}
    \label{fig:eval-latency-breakdown}
    \vspace{-2ex}
\end{figure*}

\subsection{Replay Fidelity}
\label{sec:eval-replay-valid}

We first check that \pname{} traces and replays individual LLM calls accurately by replaying each user request in isolation. 
As shown in \Cref{fig:replay-val-iso-llm}, the replayed per-LLM-call latencies closely match the original across all eight applications. 
The only visible gap is in ODR, where 4.6\% of calls exceed a 3\% relative error. 
This is because ODR issues multiple LLM calls in parallel. The serving engine cannot process all parallel calls at once, and the exact set of
queued calls during replay differs from the set during trace collection.
However, this per-call gap does not affect replay fidelity because the serving workload is determined by the dependencies among LLM calls.
We further confirm this by comparing the end-to-end user request latencies in \Cref{fig:replay-val-iso-request}.
For all applications, the replayed request latency has a mean absolute percentage error below 3\%, including ODR.
This demonstrates that \pname{} replays LLM calls with accurate timing by correctly reconstructing the execution graph of each user request.

We further verified that \pname{} can reproduce serving-system behaviors when multiple requests are served concurrently.
Recall from Figure~\ref{fig:request-timeseries-no-replay} that two independent runs of the same workload diverge significantly. We replay the first run with \pname{} for 2 hours after warmup and show the comparison in Figure~\ref{fig:request-timeseries-replay}.
With \pname{}'s replay, the serving system states matched closely throughout this observation window. 
Specifically, the mean absolute error in the number of running and waiting calls (sampled every second) is 1.4 and 1.6, respectively, while the cumulative LLM call queuing time differs by only 2.9\%.
The mean absolute percentage error of replayed requests' latency is 1.9\%. 
Together, these results confirm that \pname{} can faithfully reproduce the original application's serving-system workload, both for individual and concurrent requests, 
providing a reliable foundation for evaluating system performance.


\begin{figure*}
    \centering
    \includegraphics[trim=0 0 0 2ex,clip,width=\linewidth]{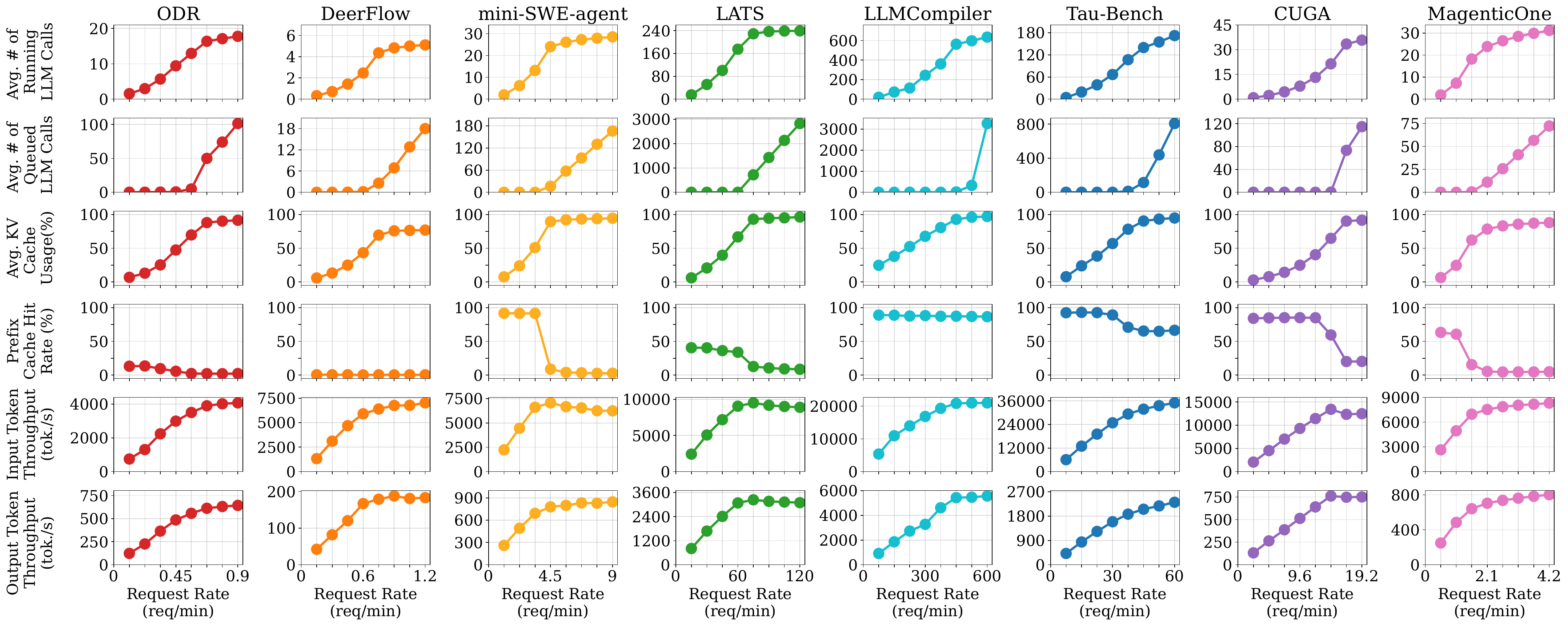}
    \vspace{-5.8ex}
    \caption{Serving system metrics.}
    \label{fig:eval-serving-system-metrics}
    \vspace{-2ex}
\end{figure*}
\begin{figure*}
    \hspace*{0.55cm}
    \includegraphics[width=\dimexpr\linewidth-0.55cm\relax]{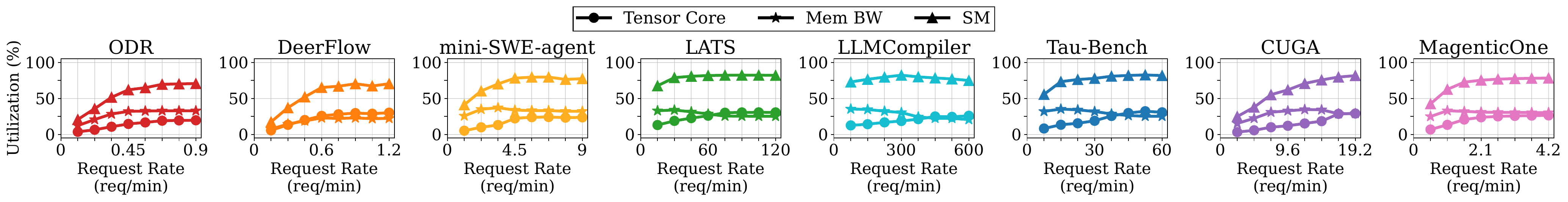}
    \vspace{-5ex}
    \caption{GPU resource utilization under different request rates.}
    \label{fig:eval-gpu-metrics}
    \vspace{-2ex}
\end{figure*}

\subsection{End-to-end Performance}
\label{sec:eval-e2e-perf}

\pname{} supports profiling the end-to-end latency of each user request and provides a detailed breakdown.
We demonstrate this by replaying collected traces and synthesizing online-serving workloads at varying request rates.


\Cref{fig:eval-latency-cdf} shows the distribution of end-to-end request latency for each application, with the request rate set below the serving system's saturation point.
The average latency varies widely across applications, as they perform different types of tasks with different difficulties.
Beyond average latency, request latency is long-tailed for most applications, with the p95 latency reaching up to 6.60$\times$ the p50 latency.
This long-tail distribution is primarily caused by agents that fail to make progress yet keep issuing LLM and tool calls instead of stopping.
DeerFlow and MagenticOne are notable exceptions, since DeerFlow enforces a strict limit on search iterations, 
while MagenticOne has an orchestrator agent that monitors other agents' progress and proactively terminates the workflow upon detecting a stall.

The two deep-research applications show a high minimum latency. Only 17\% of ODR requests complete within 175 seconds, and only 6\% of DeerFlow requests complete within 100 seconds. 
This is because a large fraction of the requests are issued by these applications as a research task, and they must complete at least one research pass and generate a long final report on the critical path.

Figure~\ref{fig:eval-latency-breakdown} provides a detailed breakdown of the end-to-end latency per output token at different request rates.
Each measurement runs for two hours. 
For applications with parallel LLM calls, we count the output tokens on the critical path.
The end-to-end latency is dominated by decoding when the serving system is not saturated, accounting for 
60\%-98\% across the studied applications.
The per-token decode latency increases with the request rate up to saturation, then stabilizes.
In contrast, prefill accounts for only a small fraction of end-to-end latency: less than 23\% for DeerFlow and 4\% on average for the remaining applications. Prefill is most significant for DeerFlow, which has the longest LLM call input prompts among the studied applications (38.8k tokens on average).
As the request rate increases, queuing delay takes a growing share of end-to-end latency, since the average queue length grows. Under heavy overloaded scenarios, queuing dominates up to 69\% of the end-to-end latency at the highest request rate we tested.

\insightbox{
    \item \pname{} enables end-to-end tracing for each user request across diverse agentic applications and provides a detailed latency breakdown.

    \item Request latency in many agentic applications is long-tailed. These tail requests arise when the agent fails to converge on a solution and issues repeated LLM calls. 

    \item At low request rates, decoding is the performance bottleneck for agentic applications. Queuing delay can become another significant source of overhead as load increases.
}

\subsection{Serving System Performance Characterization}
\label{sec:eval-sys-profile}
A key capability of \pname{} is profiling detailed metrics across the serving system and underlying hardware. Together, these metrics enable us to determine when the system saturates, its throughput at a given load level, and which resources are the limiting factors.
To demonstrate this, we profile the workloads from \S\ref{sec:eval-e2e-perf} under different request rates, reporting serving system metrics in \Cref{fig:eval-serving-system-metrics} and hardware utilization in \Cref{fig:eval-gpu-metrics}.




\noindent\textbf{Request concurrency.}
The numbers of running and queued LLM calls together indicate when the serving system saturates. As the request rate increases, the number of running calls increases until the system reaches its concurrency limit, after which this number stabilizes and any additional requests wait in the queue.
Therefore, the saturation point is the highest request rate the system can sustain while keeping the queue from growing without bound. 
For instance, with mini-SWE-agent, the number of running calls stabilizes at 24 on average, while the queue begins to grow once the request rate reaches 4.5 requests per minute.


\noindent\textbf{KV cache usage.}
At saturation, the KV cache is nearly full for most applications, making cache capacity a key factor that limits further scaling.
DeerFlow saturates at a lower usage of around 75\%, since its LLM calls have an average of 40k context tokens while our tested serving system has $\sim$213k token capacity.
At this coarse granularity, once only a small amount of cache is free, the remaining space cannot admit another call, so utilization stays low.
Also, under the pressure of increasing request rates, the prefix-cache hit rate drops, which can reduce both prefill and decode throughput.

\noindent\textbf{Prefix-cache hit rate.}
At low request rates, most applications achieve substantial prefix-cache sharing, with hit rates between 41\% and 91\%. Deep research applications have low prefix cache sharing. Specifically, ODR has a hit rate of 14\%, since its LLM call inputs mostly consist of unrelated search results. DeerFlow stays below 1\% at all request rates, since it places a timestamp at the start of its system prompt, which changes the prefix of every LLM call and eliminates nearly all prefix-cache reuse.
Removing the timestamp raises the hit rate to around 30\% at low request rates.


For the remaining applications, the hit rate drops sharply once the request rate passes a threshold and then stabilizes at a lower value as the rate increases further.
The drop reflects prefix-cache thrashing, where the serving system treats each LLM call as independent and evicts a request's accumulated context while no call is using it under KV cache pressure, even though a later call will reuse it.
For example, in mini-SWE-agent, each call reuses this context as its prefix, reaching 91\% at low rates but falling below 4\% as the rate rises. The stabilized hit rate tracks the system prompt's share of input tokens, which stays cached across requests, settling near 88\% for LLMCompiler and 61\% for Tau-Bench.

\noindent\textbf{Token throughput.}
Input and output token throughput both increase with request rate at first, then stabilize at higher rates as KV cache capacity saturates.
Added load can even reduce input-token throughput by causing prefix cache thrashing, since the serving system must recompute the evicted prefixes instead of reusing them, lowering it by up to 12\% for mini-SWE-agent.
This recomputation also interferes with decode, so output-token throughput no longer scales with the number of running LLM calls.
For instance, when mini-SWE-agent's request rate rises from 3.38 to 4.5 req/min, the number of running calls grows by 83\% while output-token throughput increases by only 11\%.

\noindent\textbf{GPU resource utilization.}
The prefix-cache thrashing is further evidenced by the GPU resource utilization.
As shown in Figure~\ref{fig:eval-gpu-metrics}, for applications that experience sharp drops in prefix-cache hit rate, GPU HBM bandwidth utilization decreases as the request rate approaches saturation.
For example, increasing the request rate for CUGA from 14.4 to 16.8 requests per minute causes tensor core utilization to increase by 55\%, while memory bandwidth utilization drops by 16\%.
During these periods, memory bandwidth utilization is lower than when the engine runs a full decode batch.
Although tensor core utilization can continue to grow at request rates beyond the saturation point, this does not indicate improved serving efficiency; rather, it reflects wasted computation due to recomputation caused by prefix-cache thrashing.

\insightbox{
    \item \pname{} collects detailed metrics from the serving system and underlying hardware, enabling users to understand how their system behaves under diverse agentic workloads at different load levels.

    \item Agent prompts should be designed with prefix-cache sharing in mind, as prompt structure significantly affects serving performance.

    \item Under KV cache pressure, an agent's prefix cache may be evicted during tool calls, causing prefix-cache thrashing and reducing the hit rate. This incurs significant recomputation costs.

}

\subsection{Support for Different Serving Scenarios}
\label{sec:eval-serving-modes}

\begin{figure}[t]
    \centering
    \includegraphics[trim=0 0 0 0.5ex,clip,width=\linewidth]{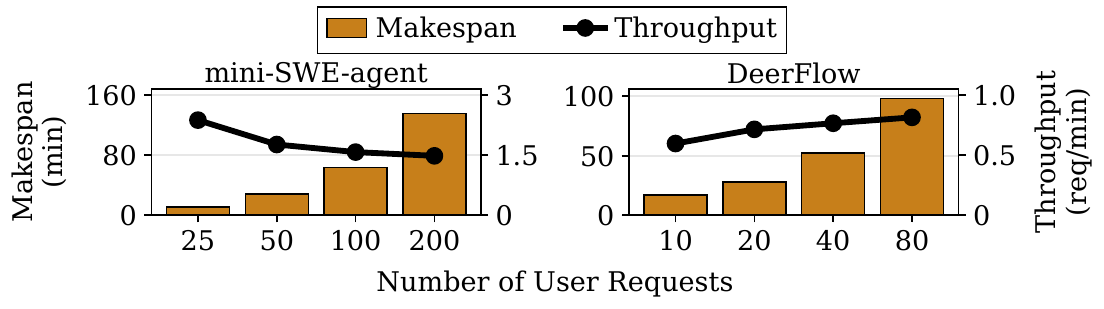}
    \vspace{-6ex}
    \caption{Offline batch inference.}
    \label{fig:load-pattern-makespan}
    \vspace{-2ex}
\end{figure}

\begin{figure}[t]
    \centering
    \includegraphics[trim=0 0 0 0.5ex,clip,width=\linewidth]{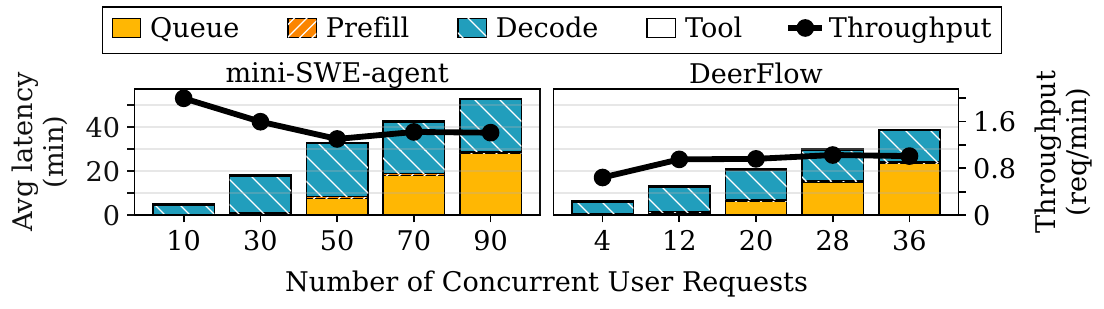}
    \vspace{-6ex}
    \caption{Fixed-concurrency serving.}
    \label{fig:load-pattern-concurrency}
    \vspace{-2ex}
\end{figure}

We have already demonstrated that \pname{} can synthesize online serving workloads with different request rates.
We now show that \pname{} can evaluate serving system performance under two additional serving scenarios.
For these experiments, we use Qwen3.6-35B-A3B.

\noindent\textbf{Offline batch inference.}
In offline batch inference setups, throughput and total processing time (i.e., makespan) are the key performance metrics.
As shown in Figure~\ref{fig:load-pattern-makespan}, mini-SWE-agent's throughput decreases by 38\% as total requests increase from 25 to 200, primarily due to prefix-cache thrashing.
DeerFlow shows the opposite trend: increasing throughput by 35\% from 10 to 80 total requests. This increase has two causes.
First, DeerFlow exposes almost no prefix-cache sharing and thus does not suffer from thrashing.
Second, a larger request count spends proportionally less time in the final drain phase where only a few long requests remain, keeping the system at high concurrency for longer.


\noindent\textbf{Fixed-concurrency serving.}
In fixed-concurrency setups, we maintain a fixed number of in-flight user requests for each application.
We break down per-request latency to study the tradeoff between concurrency and serving quality.
As shown in Figure~\ref{fig:load-pattern-concurrency}, request latency is dominated by decoding at low concurrency, but the bottleneck shifts to queuing delay as concurrency rises.
For throughput, mini-SWE-agent degrades by 35\% as concurrency increases from 10 to 50, since higher concurrency causes prefix-cache thrashing.
DeerFlow shows the opposite trend, with throughput rising from 0.64 to 0.95 req/min as concurrency increases from 4 to 36, since more concurrent requests increase the decode batch size and there is no prefix-cache to thrash.


\subsection{Scalability to Multiple Serving Engines}
\label{sec:eval-scalability}

\begin{figure}[t]
    \centering
    \includegraphics[trim=0 0 0 0.5ex,clip,width=0.9\linewidth]{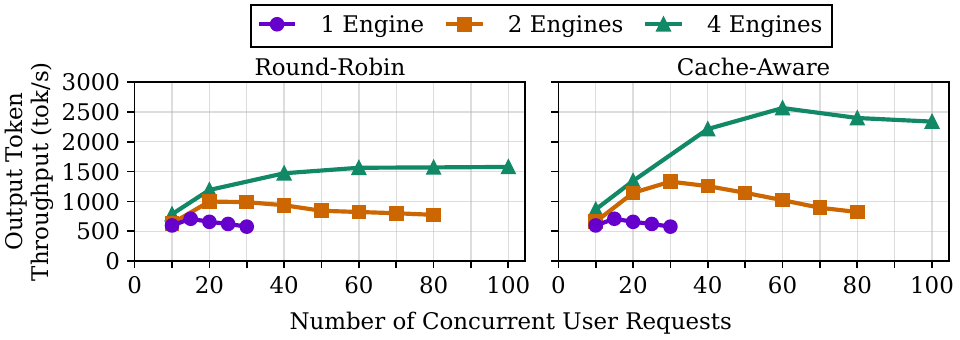}
    \vspace{-2.5ex}
    \caption{Decoding throughput as the number of serving engines scales, under different routing policies.}
    \label{fig:scalability}
    \vspace{-2ex}
\end{figure}

\pname{} can also benchmark serving systems with multiple engines across multiple nodes. 
To demonstrate this, we use Qwen3.6-35B-A3B with mini-SWE-agent to synthesize fixed-concurrency workloads. We scale the serving system from 1 to 4 engines, each holding one copy of the LLM,  distributed across two of our testbed server machines.
We compare the default round-robin and cache-aware routing policies provided by vLLM-Router. 
As shown in Figure~\ref{fig:scalability}, the serving system with round-robin policy scales poorly as we increase the number of serving engines, reaching only 1.41$\times$ and 2.26$\times$ the single-engine peak output token throughput at 2 and 4 engines, respectively. This is because with round-robin, the serving system cycles LLM calls for the same request across different engines, destroying prefix-cache locality.
In contrast, the serving system with prefix-cache-aware policy reaches 1.88$\times$ and 3.61$\times$ improvements on output token throughput at the same engine count. The policy routes almost all LLM calls from the same request to a single engine, preserving cache reuse as the system grows.
It scales decoding throughput more efficiently.


\insightbox{
    \item \pname{} can scale to multi-engine serving systems to help users evaluate distributed LLM inference systems under agentic workloads. 
    \item Realizing scale-out throughput gains depends critically on the LLM call routing policy, which must balance load across engines while exploiting prefix cache locality.
}




\subsection{Debugging Serving Systems}
\label{sec:eval-scheduling}

\begin{figure}[t]
    \centering
    \includegraphics[trim=0 0 0 0.3ex,clip,width=0.9\linewidth]{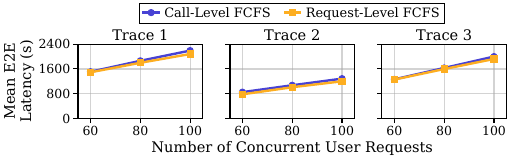}
    \vspace{-2.5ex}
    \caption{Average end-to-end latency of requests under various scheduling policies from replaying traces captured during different runs with the same random number seed.}
    \label{fig:eval-case-study-sched}
    \vspace{-2.5ex}
\end{figure}



To demonstrate how \pname{}'s trace replay help developers evaluate and debug serving system optimizations, 
we perform a case study on LLM call scheduling policies.
We compare two scheduling policies, the call-level First-Come-First-Serve (FCFS) used by mainstream serving
engines and the recently proposed request-level FCFS~\cite{infercept,li2025continuum}, which prioritizes all LLM calls belonging to an earlier user request.


We collect three traces from different runs of the mini-SWE-agent under the same fixed random seed and replay each trace under both policies.
As shown in Figure~\ref{fig:eval-case-study-sched}, with all three traces, request-level FCFS performs comparably to or slightly better (<8\%) than call-level FCFS. 
However, the results vary substantially across three traces.
Both scheduling policies achieve lower end-to-end latency (by 720 to 910\,seconds under 100 concurrent requests) on trace 2 compared to traces 1 and 3. This is because trace 2 contains fewer long requests with only 4.5\% of requests exceeding 40K total output tokens, versus 12\% and 13.5\% for traces 1 and 3.


Without \pname{}, developers may be misled due to this run-to-run variation.
If a developer happens to test call-level FCFS on the run producing trace 1 and request-level FCFS on the run producing trace 2, they would significantly overestimate the benefit of the request-level policy on this typical application. Even if they measured the call-level FCFS policy on trace 1 and the request-level FCFS policy on trace 3, the calculated benefit would be 12\%.
In contrast, \pname{}'s replay mechanism offers a fair comparison across different policies with the same trace, isolating the performance impact. They can also replay problematic runs multiple times to investigate the reason and debug the policy.

\vspace{-1ex}
\subsection{Overhead Analysis}
\label{sec:eval-overhead}
We evaluate \pname{}'s overhead by measuring its resource consumption and its interference with running workloads.
During trace recording, the OpenTelemetry tracer adds only 0.1\% to end-to-end request latency and generates 5\,KB/s of network traffic to export each request's traces over its lifetime.
The tracing backend that aggregates these traces is configured with a 1\,GB buffer for incoming traces and consumes, on average, under 5\% of a CPU core and 2\,GB of memory, writing 80\,KB/s to disk.
Logging input and output token IDs adds a further 250
\,KB/s. The trace size is 4.6\,MB per request on average, with the largest requests from DeerFlow and ODR occupying up to 27.46 and 12.3\,MB, respectively. 

During replay, \pname{} profiles detailed system and hardware metrics.
We show the overhead by comparing the average decoding iteration time with and without profiling: interference is negligible, increasing per-iteration time by <0.1\%.
Since we must fetch the metrics from the GPU and persist them to disk, keeping this traffic small is essential to avoid contending with KV cache offloading. Even at a 20\,ms sampling interval, the monitor uses only 16\,KB/s of PCIe bandwidth, negligible relative to the link speed of 64\,GB/s.


\vspace{-1.5ex}
\section{Discussion and Future Work}
\vspace{-0.3ex}

\pname{} conducts benchmarking and performance profiling of LLM serving systems under agentic workloads.
The host, which runs the application control logic and tool calls, falls outside this scope by design.
The tools of different agentic applications vary in their interfaces and execution environments.
Faithfully reproducing their execution is a distinct problem from benchmarking the serving system.

Replay remains important when the host is the object of study.
Tool calls' arguments, duration, and execution environment vary across runs, and this variance propagates into any host-side measurement taken without a fixed reference trace.
By enabling workload tracing, execution-graph construction, and faithful replay, \pname{} lays the foundation that both serving-engine and host benchmarking require.

For future work, a natural extension is to capture tool-call arguments for tools with well-defined interfaces, such as RPC calls and shell commands dispatched to Docker, where the interface boundary is explicit and the arguments are cleanly serializable.
Recording these arguments allow \pname{} to replay tool calls with their original inputs, extending faithful reproduction from the serving engine outward to the host that runs the application.

\vspace{-1.5ex}
\section{Related Work}
\vspace{-0.5ex}

\noindent\textbf{Performance Benchmarking for LLM Serving Systems.}
Existing works~\cite{llmperf_llm_inference_benchmark, reddi2020mlperf, genaiperf_llm_inference_benchmark, aiperf} evaluate LLM serving systems on independent LLM calls or multi-turn conversations, reporting metrics such as TTFT, TPOT, and decoding throughput.
More recently, AA-AgentPerf~\cite{artificialanalysis_agentperf_2026} measures serving performance with ReAct-style coding agent traces under varying numbers of concurrent sessions.
In contrast, \pname{} evaluates serving systems with diverse agentic applications, configurable load generation, detailed system-level metrics, and fully reproducible workloads.

\noindent\textbf{System Optimization for Agentic Applications.}
Many studies have been conducted to optimize serving system performance for agentic applications.
Parrot~\cite{parrot}, Ayo~\cite{ayo2025}, Kairos~\cite{kairos}, and Agentix~\cite{agentix} optimize schedule policy to reduce user request latency.
Regarding KV cache management, InferCept~\cite{infercept}, Continuum~\cite{li2025continuum}, CONCUR~\cite{concur}, and ThunderAgent~\cite{thunderagent} reduce recomputation overhead by preventing unnecessary cache evictions during tool calls, whereas KVFlow~\cite{pan2025kvflow},  PBKV~\cite{zheng2026efficient}, and ScaleSim~\cite{pan2026scalesim} proactively swap the cache to CPU memory based on predicted execution in multi-agent workflows.
Pie~\cite{gim2025pie} and AIOS~\cite{mei2025aios} ease the deployment of such optimizations by making the serving system more programmable.
\pname{} enables reproducible benchmarking of these optimizations. 

\noindent\textbf{Framework for Developing Agentic Applications.}
Early frameworks such as LangChain~\cite{langchain_sequentialchain}, LangGraph~\cite{langgraph2024}, AutoGen~\cite{wu2024autogen}, and MetaGPT~\cite{hong2024metagpt} expose APIs for building agentic workflows, but require developers to hardcode the workflow structure.
Recent frameworks~\cite{zhang2025aflow,li2024autoflow,wang2025dyflow,crewai} go further by automating the generation of the agentic workflow. 
The diversity of these frameworks makes it time-consuming to collect agentic applications for benchmarking. 
\pname{} addresses this with a unified execution graph abstraction for collecting benchmarking traces, requiring only minor code annotations to the application.

\section{Conclusion}

We develop \pname{}, a benchmarking framework that can evaluate LLM serving systems under diverse agentic workloads.
\pname{} provides an end-to-end pipeline to collect, synthesize, and replay agentic workloads with minimal effort required from users.
Our experiments demonstrate that \pname{} can accurately replay agentic workloads while capturing detailed system metrics at low overhead.


\bibliographystyle{ACM-Reference-Format}
\bibliography{ref}

\end{document}